\documentclass[sigconf]{acmart}
\AtBeginDocument{%
  }
\usepackage{xspace}
\setcopyright{acmlicensed}
\copyrightyear{2018}
\acmYear{2018}
\acmDOI{XXXXXXX.XXXXXXX}
\acmConference[Conference acronym 'XX]{Make sure to enter the correct
  conference title from your rights confirmation email}{June 03--05,
  2018}{Woodstock, NY}
\acmISBN{978-1-4503-XXXX-X/2018/06}

\newcommand{\ours}{\textsc{NgCaptcha}\xspace}

\begin{document}

\title{\ours: A CAPTCHA Bridging the Past and the Future}

\author{Ziqi Ding}
\email{ziqi.ding1@unsw.edu.au}
\orcid{0009-0007-6257-1502}
\affiliation{%
  \institution{UNSW Sydney}
  \city{Sydney}
  \state{New South Wales}
  \country{Australia}
}

\author{Shangzhi Xu}
\email{shangzhi.xu@unsw.edu.au}
\orcid{}
\affiliation{%
  \institution{UNSW Canberra}
  \city{Canberra}
  \country{New South Wales}}
\email{Australia}

\author{Wei Song}
\email{wei.song1@unsw.edu.au}
\orcid{}
\affiliation{%
  \institution{UNSW Sydney}
  \city{Sydney}
  \country{New South Wales}}
\email{Australia}

\author{Yuekang Li}
\email{yuekang.li@unsw.edu.au}
\orcid{}
\affiliation{%
  \institution{UNSW Sydney}
  \city{Sydney}
  \country{New South Wales}}
\email{Australia}

\renewcommand{\shortauthors}{Trovato et al.}

\begin{abstract}
CAPTCHAs are widely employed for distinguishing humans from automated bots online.  
However, 
current vision-based CAPTCHAs face escalating security risks:
traditional attacks continue to bypass many deployed CAPTCHA schemes, 
and recent breakthrough in AI, 
particularly in large-scale vision models, 
enable machine solvers to significantly outperform humans on many CAPTCHA tasks, 
undermining their design assumptions. 
To address these issues, 
we introduce \ours, a \textbf{N}ext-\textbf{G}eneration \textbf{CAPTCHA} framework that integrates a lightweight client-side proof-of-work (PoW) mechanism with an AI-resistant visual recognition challenge.
In \ours, a browser must first complete a small hash-based PoW before any challenge is displayed, throttling large-scale automated attempts by increasing their computational cost.
Once the PoW is solved, the user is presented with a human-friendly but model-resistant image selection task that exploits perceptual cues current vision systems struggle with.
This hybrid design couples computational friction with AI-robust visual discrimination, substantially raising the bar for automated bots while keeping the verification process fast and effortless for legitimate users.



\end{abstract}

\ccsdesc[500]{Security and privacy~Access control}

\keywords{CAPTCHA, Proof-of-work, Illusion}


\maketitle

\section{Introduction}

CAPTCHAs have long served as a foundational defense against automated abuse on the web, distinguishing human users from bots through tasks that are intuitive for people yet difficult for machines \cite{ding2025illusioncaptcha}.
Since the term CAPTCHA was introduced in 2003 by von Ahn et al.~\cite{von2003captcha}, 
widespread deployments have included distorted-text transcription~\cite{wang2023experimental}, object-based image selection~\cite{madathil2019empirical}, and other perception-oriented challenges \cite{ding2025illusioncaptcha}.

However, traditional CAPTCHA mechanisms are increasingly failing in practice.
Rapid advances in artificial intelligence, particularly large vision–language models, now enable automated systems to solve both text-based and image-based CAPTCHAs with high reliability~\cite{deng2024oedipus}.
Recent studies~\cite{deng2024oedipus,teoh2025captchas,song2025help} demonstrate that state-of-the-art algorithms can consistently break many of the visual CAPTCHAs deployed by major websites.
In parallel, human-solver services (``CAPTCHA farms'') have reduced the economic barrier for adversaries: outsourcing CAPTCHA solving costs as little as \$0.50 per 1,000 challenges~\cite{ding2025illusioncaptcha}.
Together, these developments render existing CAPTCHA schemes ineffective against both automated and human-assisted attacks.
Beyond security limitations, CAPTCHAs also introduce usability and privacy concerns.
User friction, such as deciphering distorted text or selecting objects across multiple images can degrade the browsing experience, while widely used services like Google reCAPTCHA raise privacy issues due to their reliance on behavioral tracking for risk assessment.

To address these challenges, researchers have explored \textit{proof-of-work (PoW) CAPTCHAs}, 
which shift the burden of verification from human cognition to client-side computation.
Instead of solving a visual or linguistic puzzle, the user's device computes a small cryptographic hash puzzle in the background, typically requiring only a few seconds of CPU time \cite{chadam2023proof}.
This design effectively blocks traditional bots by imposing direct computational costs while keeping the verification process invisible to human users.
Commercial systems such as Friendly Captcha \cite{friendlycaptchaFriendlyCaptcha} adopt this paradigm, replacing interactive challenges with lightweight PoW tasks that run automatically on the client.
Such approaches improve accessibility, 
reduce user friction, 
and avoid the privacy concerns associated with third-party CAPTCHA providers, since the mechanism can be fully self-hosted without collecting behavioral 
or personal data.
However, 
PoW alone does not reliably distinguish humans from AI agents.
A sufficiently resourced adversary, 
e.g., leveraging GPUs, cloud infrastructure, or distributed botnets, can still solve PoW puzzles at scale.
Thus, 
while PoW raises the economic cost of abuse, 
it cannot prevent determined attackers from automating verification.

In this paper, we present \ours, a hybrid system that combines the strengths of both paradigms: it couples a lightweight client-side PoW puzzle with a AI-hard image recognition challenge. The insight is that by requiring a small PoW before presenting the CAPTCHA question, automated bots are forced to expend non-trivial computation for each attempt, dramatically slowing them down or discouraging them, while legitimate human users experience only a minor delay. After the PoW is solved, the user is prompted to select illusionary images, a task that remains easy for humans but still non-trivial for machine vision in many cases~\cite{ding2025illusioncaptcha}. Our approach thus raises the bar for attackers in two dimensions (CPU and GPU) while keeping the overall user experience relatively seamless. In our prototype implementation, the entire process typically adds only a couple of seconds of background computation before a user can quickly complete a familiar image selection task.

We have developed a fully functional demo of \ours{} as a self-contained web application. In the following sections, we describe the system design (Section~\ref{sec:methods}), evaluate its performance and security characteristics (Section~\ref{sec:results}), and discuss its advantages, limitations, and potential improvements (Section~\ref{sec:discussion}). This work illustrates a new direction for CAPTCHA technology: leveraging client-side computational puzzles alongside human-centric visual questions to achieve more robust bot resistance.

\section{Methods and System Design}\label{sec:methods}

\ours{} operates in two sequential stages: a client-side proof-of-work (PoW) computation followed by an AI-resistant image selection challenge. 
Figure~1 illustrates the overall workflow. 
For our prototype, 
we built the server in Python using a Flask backend and implemented the client interactions through a lightweight JavaScript front-end. 
The following sections describe each stage of the pipeline and the corresponding system components in detail.

\begin{figure}[htbp]
    \centering
\includegraphics[width=\linewidth]{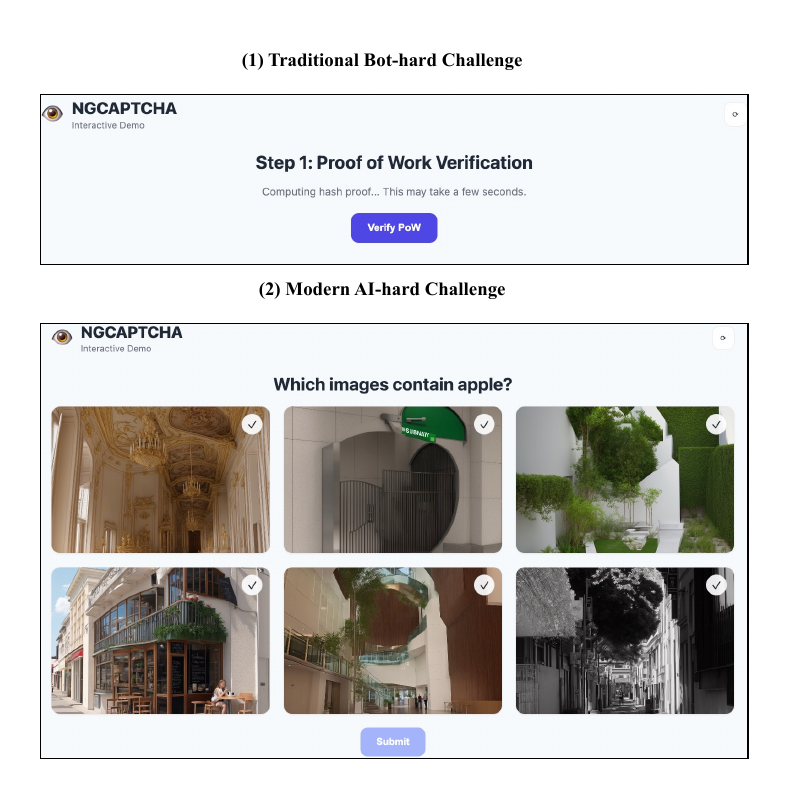}
    \caption{Design of \ours.}
    \label{fig:NGCAPTCHA}
\end{figure}

\subsection{Phase 1: Proof-of-Work Challenge}
When a client initiates a CAPTCHA verification request, 
the server first issues a proof-of-work challenge rather than immediately presenting a question. 
To construct this challenge, 
the server generates a random cryptographic seed (e.g., a nonce) and specifies a target difficulty level that determines the required amount of client-side computation.
In our implementation, 
the proof-of-work challenge consists of a randomly generated salt value and a target difficulty parameter~$d$.
The client must compute a numeric nonce such that the SHA-256 hash of the concatenation of the salt and nonce contains at least~$d$ leading zero bits, mirroring the structure of standard hash-based PoW puzzles.
This construction is analogous to Hashcash-style or blockchain mining puzzles, 
but with a difficulty level calibrated to be solvable within a few seconds on commodity devices.
For example, if $d=4$ (the default in our demo), the hash must start with 4 zero hex digits, 
which corresponds to $4 \times 4 = 16$ leading zero bits – roughly a 1 in $2^{16}$ chance for any given nonce.

The PoW computation is executed client-side in JavaScript.
To avoid blocking the main UI thread, we offload the hashing loop to a dedicated Web Worker, which iteratively computes SHA-256 digests while incrementing the nonce until the difficulty condition is satisfied.
Once a nonce satisfying the difficulty requirement is found, the client submits the solution to the server via the \texttt{/api/pow-verify} endpoint.
The server verifies the proof-of-work efficiently by recomputing the SHA-256 digest from the provided salt and nonce and checking whether the result meets the required zero-bit prefix constraint.
If the submitted PoW solution is valid and received within the permitted time window (which prevents replay of stale challenges), 
the server accepts that the client has ``paid'' the required computational cost and proceeds to issue the actual CAPTCHA task.
This verification step is lightweight, 
consisting only of a single hash recomputation so it imposes negligible overhead on the server and does not contribute to backend load.
Notably, until the PoW is solved correctly, the server does not send any CAPTCHA images or questions — thwarting trivial denial-of-service attacks where bots could otherwise mass-request expensive CAPTCHA generation. The proof-of-work functions as a gatekeeper: only clients that put in the required computation get to attempt the visual challenge.

Each PoW challenge is tied to a session token to prevent sharing or reusing solutions across different clients. The server invalidates the PoW token after one successful verification or after a timeout to prevent replay attacks. The difficulty level $d$ can be configured; higher values increase the required computation exponentially. For demonstration purposes, we set $d=4$ which is suitable for desktops and modern smartphones. To further defend against malicious traditional bot attacks, we can set $d=5$, which increases the required computation time.

\subsection{Phase 2: Image Selection Challenge}
After the client passes the PoW hurdle, the server responds with an AI-hard image-based CAPTCHA challenge. Our system maintains a collection of images organized into several semantic categories. In this part, we leverage the concept from IllusionCAPTCHA~\cite{ding2025illusioncaptcha}, an AI-hard CAPTCHA known for exploiting perceptual illusions to make it difficult for machine learning models to solve. Unlike the original design, which prompts human users to select description options for images, our approach draws inspiration from Google’s reCAPTCHA system. Specifically, we use a grid-based image selection format, which makes it easier for human users to choose the correct images based on perceptual cues while adding an additional layer of challenge for AI systems.

By utilizing this hybrid design, we simplify the user interaction by enabling a familiar and intuitive process, similar to reCAPTCHA’s image grid selection. This familiarity makes it easier for humans to identify objects, as they can rely on basic visual cues. However, for AI models, the challenge is significantly more complex. While AI systems excel at recognizing specific patterns or objects, the perceptual illusions embedded in the images make it much harder for them to correctly categorize the images, especially as the illusions cause the content to appear ambiguous or distorted.

\section{Results and Preliminary Evaluation}\label{sec:results}


We conduct comprehensive evaluations of \ours, covering three key perspectives: construction cost, user safety, and usability.

\smallskip
\noindent
\textbf{Construction Cost.} On the server side, the performance overhead of \ours{} is very low. Proof-of-work verification requires only a single hash computation plus a few table lookups, and returning a set of six images is negligible work for a standard web server. The memory footprint is also modest: the server only maintains small records for active sessions/challenges, while static images are loaded from disk on demand. Even under high request rates, the PoW requirement naturally throttles abusive traffic. A malicious client attempting to request challenges repeatedly must solve each PoW instance, making it difficult to flood the server with valid requests. In our implementation, we observed no noticeable performance degradation with multiple concurrent users, indicating that \ours{} scales well for typical deployment scenarios. Overall, the prototype operates correctly and efficiently: legitimate users solve the CAPTCHA quickly, while automated attempts that do not solve the PoW are effectively suppressed.

\smallskip
\noindent
\textbf{Security Test.} The additional computational cost significantly increases the effort required for an automated attacker. With a difficulty level of $d=4$, the nonce search space is approximately $2^{16}$ (65,536). While an attacker controlling a botnet or a powerful machine could accelerate the search via parallelism or hardware optimization, doing so still requires a substantial investment of resources. Even if a botnet were able to solve a challenge in, say, $0.5$ seconds on average (by leveraging multiple CPU cores or GPUs), the computational burden remains considerable when scaled to a large campaign. For instance, to solve 100,000 CAPTCHAs (as might be required in a large-scale flooding attack on a website's signup form), the attacker would need to perform on the order of $10^{11}$ hash computations. In contrast, a traditional CAPTCHA attack could outsource 100,000 challenges to human solvers for a relatively small fee, or utilize a machine learning model to solve them almost instantaneously, if such a model exists. With the introduction of \ours{}'s PoW mechanism, however, an attacker must expend nontrivial computational resources for each individual attempt. This fundamentally disrupts the economics of large-scale attacks: solving 100,000 challenges would require several days of aggregated CPU/GPU time, vastly increasing the attack cost and making such campaigns far less feasible. Additionally, the use of locally stored image categories ensures that the exact content of the CAPTCHA challenges is not known to the attacker in advance; the images can only be downloaded and analyzed after a valid PoW solution has been submitted. In our experiments, the server reliably blocked automated scripts that did not complete the PoW step: any attempt to fetch the image challenge without providing a valid PoW solution was denied. Only after successfully solving the PoW does the server release the corresponding image challenge, ensuring that malicious clients cannot bypass the computational puzzle.

Another important security consideration is that our current illusionary image challenges are intentionally designed to be easy for human users. It is conceivable that a sufficiently advanced computer vision model could also recognize these objects with high accuracy, thereby partially automating the second step. However, existing models still struggle with illusionary images~\cite{ding2025illusioncaptcha}, which means our image-selection challenge is likely to remain effective against AI-based attacks for a considerable time. Moreover, even if a bot attempts to leverage AI to solve the images, it must rely on a highly capable vision model, as current vision and multi-modal systems (including GPT-based agent frameworks) still find it difficult to robustly handle our specific image-selection tasks, as demonstrated in Table~1. 

\begin{table}[t]
  \centering
  \caption{Performance of different GPT models under Zero-shot and Chain-of-Thought prompting to identify Image Selection Challenges.}
  \label{tab:gpt-zero-cot}
  \begin{tabular}{lcc}
    \toprule
    Model & Zero-shot & Chain-of-Thought \\
    \midrule
    GPT-5.1      & 0.00\% & 0.00\%\\
    GPT-4o       & 0.00\% & 0.00\% \\
    GPT-4o-mini  & 0.00\% & 0.00\% \\
    \bottomrule
  \end{tabular}
\end{table}

Moreover, the combination of two different challenge types (hash puzzle + image recognition) forces an attacker to succeed in both domains. This defense-in-depth design makes it substantially harder to build a single automated solver that is simultaneously efficient at cryptographic PoW and semantic image understanding. In our tests, we also simulated an attack scenario in which a script first solves the PoW and then randomly selects images (i.e., guessing without actually analyzing the content). As expected, such random guessing never succeeded in a single attempt due to the very low probability of randomly choosing the correct 2 or 3 images out of 6. The attacker would have to repeatedly pay the PoW cost for many guesses, rendering this approach impractical. This underscores how the PoW requirement compels the attacker to ``pay as they go'' for each guess, whereas a human user typically solves the CAPTCHA correctly in one attempt.

\smallskip
\noindent
\textbf{Usability Test.} In the default configuration, the proof-of-work puzzle is typically solved by the browser in approximately 1-2 seconds. This overhead is small: it adds only a brief delay before the user sees the image selection prompt. In practice, users reported that the transition was almost unnoticeable or only a minor wait. The image selection phase then proceeds like a normal visual CAPTCHA; users(n=10) in our informal tests were able to identify and click the correct images usually within 3–5 seconds. All human testers succeeded on the first attempt for each challenge, finding the tasks straightforward. Thus, the total time to solve \ours{} averaged around 5-7 seconds, which is on par with or even faster than many traditional CAPTCHAs that often take over 10 seconds for users to complete. Notably, the users do not need to solve any distorted text or puzzle themselves; aside from a quick click task, the only ``work'' they do is letting their device compute for a moment. This suggests the hybrid approach remains user-friendly. Furthermore, we tested our PoW verification on various smartphones, and the results show that it can be completed quickly across a wide range of devices.

\section{Demonstration Overview}\label{sec:discussion}
Our \ours{} demonstrates a viable path toward more secure and user-friendly anti-bot measures, but it also raises several points for discussion and future improvement.

\smallskip
\noindent \textbf{Advantages.} The primary advantage of combining PoW with a traditional CAPTCHA is the significant increase in attack cost. A spam bot is no longer limited only by the cleverness of its AI solver or the cheapness of hired humans, but also by raw computational constraints. This economic deterrent makes large-scale abuse much less attractive. Even if an attacker uses human solvers, they would need to pay not just for solving the images but also bear the computation cost (or rent botnet time) for the PoW, which is not something that can be easily outsourced for free. Our system also provides privacy benefits: it does not rely on any external service or collect extensive user behavior data, unlike many existing CAPTCHA solutions. Everything runs under the website owner's control, which is appealing for GDPR compliance and for deployment in sensitive applications. Additionally, the user experience in our approach can be quite seamless—aside from the quick image selection, the process can feel nearly invisible if the PoW is solved in the background with minimal delay. This is in line with trends to reduce friction for legitimate users.

\smallskip
\noindent \textbf{Limitations.}  Despite its advantages, we acknowledge that \ours{} is not a panacea. Determined attackers with substantial resources, specialized hardware, or large botnets could still amortize the proof-of-work cost, since PoW CAPTCHAs can only make attacks more expensive, not impossible. Likewise, the image-recognition step may be weakened as vision and multi-modal models improve, especially if the challenge space is small or static. This calls for diverse, regularly updated, and possibly illusion-based images to help maintain a human–AI gap.

Our current design also inherits the accessibility limitations of typical image CAPTCHAs: visually impaired users cannot complete the challenge. In practice, a deployment should offer alternative modalities, such as audio CAPTCHAs or a purely PoW-based fallback that requires no visual interaction but more computation. Providing a clear, user-friendly way to choose or trigger these alternatives is essential for real-world adoption and accessibility compliance. Therefore, it is important to develop a more general CAPTCHA that works for all human users. Also, we need to update CAPTCHA steadily as the evolvement of AIs is very fast.

\smallskip
\noindent \textbf{Future Work.} Moving forward, we plan to evaluate NGCAPTCHA with a larger user study and under simulated attack conditions. User feedback on the perceived delay and any confusion during the process will be valuable. We expect that many users will not even realize a computational puzzle was solved by their browser (especially if we add a small message like “Securing...please wait” during that phase). We are also interested in quantifying the security more formally: e.g., measuring how increasing the PoW difficulty impacts automated attack success rates, and at what point the cost becomes prohibitive. Integrating more complex visual challenges, such as those based on adversarial examples or illusions, is another avenue to explore for staying ahead of AI solvers. Finally, exploring alternate proof-of-work algorithms (such as memory-hard functions or verifiable delay functions) could further improve resistance against specialized hardware and provide adjustable puzzles that are fair across different device capabilities.

\smallskip
\noindent \textbf{Conclusion.}In this work we introduced \ours{}, a hybrid CAPTCHA that combines a client-side proof-of-work puzzle with an illusion-based image selection task. Requiring a small amount of local computation before any visual challenge is revealed raises the economic bar for large-scale automated abuse while keeping the cost acceptable for legitimate users. The illusionary image grid further exploits perceptual cues that remain difficult for current vision and multi-modal models, preserving a practical gap between human and machine performance. Our prototype indicates that this defense-in-depth design can be deployed with low server overhead, only a few seconds of additional latency, and good scalability under higher request rates. Although powerful attackers, future AI advances, and accessibility concerns remain important challenges, \ours{} illustrates a promising direction for bot defense: combining client-side computational friction with human-centric, AI-resistant visual tasks to deliver more robust and privacy-preserving protection on the web.
\bibliographystyle{ACM-Reference-Format}
\bibliography{sample-base}

\appendix
\end{document}